\documentclass[aps,pre,twocolumn,superscriptaddress,secnumarabic,showpacs,floatfix]{revtex4}
\usepackage{latexsym}
\usepackage{amsmath}
\usepackage{amssymb}
\usepackage{amsfonts}
\usepackage{graphicx}

\begin{document}

\title{Anomalous diffusive behavior of a harmonic oscillator \\
driven by a Mittag-Leffler  noise}

\author{A. D. Vi\~{n}ales}

\affiliation{Departamento de F\'{\i}sica, Facultad de Ciencias
Exactas y Naturales,\\ Universidad de Buenos Aires, 1428 Buenos
Aires, Argentina.}


\author{K.  G. Wang}


\affiliation{Physics and Space Science Department, \\
Florida Institute of Technology, Melbourne, FL 32901-6975, USA\\}


\author{M. A. Desp\'osito}

\email[]{mad@df.uba.ar}

\affiliation{Departamento de F\'{\i}sica, Facultad de Ciencias
Exactas y Naturales,\\ Universidad de Buenos Aires, 1428 Buenos
Aires, Argentina.}

\affiliation{Consejo Nacional de Investigaciones Cient\'{\i}ficas
y T\'{e}cnicas, Argentina.}


\begin{abstract}

The diffusive behavior of a  harmonic oscillator driven by a Mittag-Leffler  noise is studied.
Using Laplace analysis we derive exact expressions for the relaxation functions of the particle
in terms of generalized Mittag-Leffler functions and its derivatives from a generalized Langevin equation.
Our results show that the oscillator displays an anomalous diffusive behavior.
In the strictly asymptotic limit, the dynamics of the harmonic oscillator
corresponds to an oscillator driven by a noise with a pure power-law autocorrelation function.
However, at short and intermediate times the dynamics has qualitative difference due to the presence
of the characteristic time of the noise.

\end{abstract}



\maketitle

\section{Introduction}

The study of anomalous diffusion in complex or
disordered media  has achieved a substantial progress during the last  years
\cite{Wa1,Bu,MeK,Bar1,BarSil,Die,Yil}.
Anomalous diffusion in physical and biological systems
can be formulated in the framework of the generalized Langevin equation (GLE) \cite{Wa1,Porr,Wang3,Wa,Chau,ChaKou,Kou,Fa,Fa1,Yil}.
If one  considers the dynamics  of a harmonic oscillator with frequency $\omega$
under  the influence of a random force
modeled as Gaussian colored noise,
the corresponding GLE is written as \cite{Wang3,Wa,VD}:
\begin{eqnarray}
\ddot{X}(t) + \int_0^t dt' \gamma (t-t') \dot{X}(t') + \omega^2 X
= \xi(t) \, , \label{Lang}
\end{eqnarray}
where   $X(t)$ represents the position of a particle of mass $m = 1$ at
 time $t$,
and $\gamma(t)$ is the frictional memory kernel.  The random force $ \xi(t) $ is
zero-centered and stationary Gaussian  that obeys the
fluctuation-dissipation theorem \cite{Zwa}
\begin{eqnarray}
\langle \xi(t) \xi(t') \rangle = C(|t - t'|) =k_B T \, \gamma(|t -
t'|)
 \, ,
\label{tfd}
\end{eqnarray}
where $ k_B $ is the Boltzmann constant, and  $T$ is the absolute temperature
of the environment.


It is now well established that the physical origin of anomalous diffusion
is related to the long-time tail correlations \cite{Wa1,MeK}.
Therefore, in order to model anomalous diffusion  process,
pure power-law correlation functions are  usually employed
\cite{Wa1,Porr,VD,Wa,Lut,Bark}, which may be written as
\begin{eqnarray}
C(t)=  C_{\lambda} \, \frac{|t|^{- \lambda}}{\Gamma( 1 -\lambda)}  \, ,
\label{memdespl}
\end{eqnarray}
where  $\Gamma(z) $ is the Gamma function, and $C_{\lambda}$ is a proportionality
coefficient dependent on the
exponent $\lambda$ but independent of time. The exponent $\lambda$
can be taken as $ 0 < \lambda < 1 $ or $ 1 < \lambda < 2 $, which is determined
by the dynamical mechanism of the physical process considered.

Vi\~{n}ales and Desp\'osito have introduced a  noise whose correlation is
proportional to a Mittag-Leffler function \cite{VD1}.
This correlation  behaves as a power-law for large times,  but is non-singular at the origin
due to the inclusion of a characteristic time.

The aim of this  work is to  investigate the effects of the Mittag-Leffler noise  on the behavior of a
harmonically bounded particle governed by the GLE (\ref{Lang}).
This paper is organized as follows. In Section 2 we discuss some  characteristics of the  Mittag-Leffler  noise.
In Section 3, we show the formal expressions for the relaxation functions that govern the
dynamics  of the particle in the case of an arbitrary noise correlation function.
Analytical solutions of the GLE for a harmonically bounded particle driven
by a Mittag-Leffler noise are obtained in Section 4.
The Section 5 is devoted to the analysis of
temporal behavior of  the relaxation functions, and  is compared with the results in the case of a
pure power-law noise correlation function.
Finally, the conclusions are presented in Section 6.

\section{ Mittag-Leffler  noise}

It is well known that if the correlation function (\ref{tfd}) is
a Dirac delta function the stochastic process is Markovian and its
dynamics can be directly obtained \cite{Ris}.
However, in a complex or viscoelastic environment one must take
into account the memory effects through a long-time tail noise to describe the effect of
the environment on the particle. The non-Markovian
dynamics is involved in these physical processes.

 Recently, Vi\~{n}ales and Desp\'osito introduced a Mittag-Leffler noise given by  \cite{VD1}
\begin{eqnarray}
C(t)=  \frac{C_{\lambda}}{\tau^{ \lambda}} E_{\lambda} (-( |t|/\tau
)^{\lambda}) \, , \label{memdes}
\end{eqnarray}
where $\tau$ acts as a characteristic memory time and  $ 0 < \lambda < 2 $.
The $E_{\alpha} (y)$ function  denotes the Mittag-Leffler function
\cite{Er} defined through the following series
\begin{eqnarray}
E_{\alpha}(y) = \sum_{j = 0}^{\infty} \frac{y^{j}}{\Gamma( \alpha
j + 1)}   , \quad \alpha > 0 \,  . \label{mittaglef1p}
\end{eqnarray}

Using the asymptotic behaviors of the Mittag-Leffler function
\cite{Go} one  can easily deduce that, for $ \lambda \neq 1 $, the
correlation function (\ref{memdes}) behaves as a stretched
exponential for short times and as an inverse power law in the
long time regime \cite{Go,MeGr}.

Setting $ \lambda = 1 $,
the correlation function  (\ref{memdes}) reduces to an exponential
form
\begin{eqnarray}
C(t)=  \frac{C_{1}}{\tau} \, e^{-|t|/\tau}  \, ,
\label{memdesex}
\end{eqnarray}
which describes a standard Ornstein-Uhlenbeck process
\cite{Ris}. Moreover, in the limit $ \tau \to 0 $ and from the limit representation  of the
Dirac delta \cite{Cohen} we get that $C(t)=  2 \, C_1 \,\delta (t)$
which corresponds to a white noise, non-retarded friction and
standard Brownian motion \cite{Ris}.

On the other hand, for $\lambda\neq 1$ the limit $ \tau \to 0$  of the proposed correlation
function (\ref{memdes}) reproduce the power-law correlation function (\ref{memdespl}).
This behavior is  obtained introducing in expression
(\ref{memdes}) the asymptotic behavior at large $y$ of the
Mittag-Leffler function \cite{Go}
\begin{eqnarray}
E_{\alpha}(-y) \thicksim  [y \, \Gamma( 1- \alpha)] ^{-1}  , \quad
y > 0  \, .
 \label{mitag1pas}
\end{eqnarray}

It is worth pointing out that the Mittag-Leffler correlation function  (\ref{memdes})
is a well defined and no-singular function.
From (\ref{memdes}), its value at $t=0$ is given by
$C(0)=  C_{\lambda}/\tau^{ \lambda}$,
while  for the power-law correlation (\ref{memdespl})
$C(0)$ diverges.
Then, the introduction of the characteristic time $\tau$ enables
to avoid the singularity of the power-law at the origin.
Considering that the Mittag-Leffler function is
the natural generalization of the exponential function \cite{Er},
we can also consider the Mittag-Leffler correlation function as
a generalization of the power-law correlation, and similarly,
the colored noise (\ref{memdesex}) is considered as a generalization of the white noise.

\section{Solutions of the generalized Langevin equation}

In what follows we consider the Langevin equation (\ref{Lang})
with the deterministic initial conditions $x_0 = X(0)$ and $v_0 = \dot{X}(0)$.
By means of the Laplace transformation to Eq. (\ref{Lang}) one can easily obtain a
formal expression for the displacement $ X(t)$ and the velocity
$V(t)= \dot{X}(t)$. The displacement $ X(t)$ satisfies that
\begin{eqnarray}
X(t) =  \langle X(t) \rangle  + \int_0^t dt' G(t-t') \xi(t') \, ,
\label{X}
\end{eqnarray}
where
\begin{eqnarray}
 \langle X(t) \rangle =   v_{0} \, G(t) + x_0 (1 - \omega ^2  I(t)) \,
 \label{VMX}
\end{eqnarray}
is the position mean value.
The  relaxation function  $G(t)$ is the  Laplace inversion of
\begin{eqnarray}
\widehat{ G}(s) =  \frac{1}{ s^2 + \widehat{\gamma}(s) s + \omega ^2
} \, ,
 \label{kernella}
\end{eqnarray}
where $\widehat{\gamma}(s)$ is the Laplace transform of the
damping kernel and
\begin{eqnarray}
 I(t) =  \int_0^t dt' G(t')\, .
 \label{relaIG}
\end{eqnarray}

On the other hand, the velocity $V(t) $  satisfies that
\begin{eqnarray}
V(t) =  \langle V(t) \rangle  + \int_0^t dt' g(t-t') \xi(t') \, ,
\label{V}
\end{eqnarray}
where
\begin{eqnarray}
 \langle V(t) \rangle =   v_0 \, g(t) - \omega ^2 \, x_0 \, G(t) \, ,
 \label{VMV}
\end{eqnarray}
is the velocity mean value and the relaxation function $g(t)$ is the derivative of $G(t)$, i.e.
\begin{eqnarray}
  g(t) =  G'(t)\, .
  \label{relaGg}
\end{eqnarray}

Explicit expressions of the variances can be obtained
from Eqs.(\ref{X}) and (\ref{V}). Taking
into account the symmetry property of the correlation function and
Eq.(\ref{tfd}), yields \cite{Wang3,Wa,VD,Fa1}
\begin{eqnarray}
\beta \,\sigma_{xx}(t) &=&  2 \, I(t) - G^{2}(t) -\omega^2
I^2(t) \, , \label{sigmaxni}\\
\beta \,\sigma_{vv}(t) &=& 1 - g^{2}(t) -\omega^2 G^{2}(t) \, ,
\label{sigmavni}\\
\beta \,\sigma_{xv}(t) & = & G(t)\left \{ 1 - g(t) -\omega^2
I(t)\right \} \, ,  \label{sigmaxvni}
\end{eqnarray}
where $ \beta =  1 / k_B T  $.

From an experimental point of view,  information about the
observed diffusive behavior is extracted from the mean square
displacement $\rho(t)$.
In the long time measurement time,
it is related to the relaxation function $I(t)$ as \cite{DV2}
\begin{eqnarray}
\rho(\tau_L)=  \lim_{ t \to \infty } \langle \left(X(t+
\tau_L  )-X( t )\right)^{2} \rangle =2 k_B T I(\tau_L)
\label{msd} \, ,
\end{eqnarray}
where $\tau_L$ is the so-called time lag.
Alternative information about the dynamics  can be
extracted from the normalized velocity autocorrelation function $C_V(t)$,
which  is related to the relaxation
function $g(t)$ as \cite{VD,DV2}
\begin{eqnarray}
C_V(\tau_L) = \lim_{ t \to \infty} \frac{\langle
V( t + \tau_L )V(t)\rangle} {\langle
V(t )V(t)\rangle} = g(\tau_L) \, .
\label{cv}
\end{eqnarray}

Then, the knowledge of the relaxation functions $I(t)$, $G(t)$ and $g(t)$
allows us to describe the diffusive behavior of the oscillator.
In the next section we will give explicit expressions for the relaxation
functions in the case of a Mittag-Leffler noise (\ref{memdes}) assuming that $\lambda\neq 1$.

\section{ Analytical relaxation functions for a Mittag-Leffler  noise}

From relation (\ref{tfd}),
 the memory kernel $\gamma(t)$ corresponding to the Mittag-Leffler noise
 (\ref{memdes})  can be written as
\begin{eqnarray}
\gamma(t)=  \frac{\gamma_{\lambda}}{\tau^{ \lambda}} E_{\lambda} (-( |t|/\tau
)^{\lambda})  \, ,
\label{memdes2}
\end{eqnarray}
where $\gamma_{\lambda} = C_{\lambda}/k_B T$ .
%
%
Taking into account that the  Laplace transform of the memory kernel reads \cite{Go}
\begin{eqnarray}
\label{memdesla} \widehat{\gamma}(s) = \frac{ \gamma_\lambda s^{ \lambda -1 }}
{1 + {s}^{ \lambda}{\tau}^{ \lambda}} \, ,
\end{eqnarray}
the relaxation function $I(t)$ can be written as  the Laplace inversion of
\begin{eqnarray}
\widehat{I}(s) =  \frac{\widehat{ G}(s)}{ s} = \widehat{ I_0}(s)
+ \widehat{I_1}(s)  \label{kernellades} \, ,
\end{eqnarray}
where
 \begin{eqnarray}
 \widehat{I_0}(s) & = &\frac{s^{-1}}{ {\tau}^{
 \lambda}  s^{2+ \lambda} +
  s ^{ 2 } + \bar{\gamma}_{\lambda} s ^{\lambda} + {\omega}^2}
 \label{kernellades0} \, ,
 \\
 \widehat{I_1}(s)  &=& {\tau}^{ \lambda} s^{ \lambda} \widehat{
 I_0}(s) \label{kernellades1} \, ,
 \end{eqnarray}
and $ \bar{\gamma}_{\lambda}$ is defined as
 \begin{eqnarray}
 \bar{\gamma}_{\lambda} = \gamma_{\lambda} +   \omega^2 {\tau}^{\lambda} \,  .
 \label{gammanuevo}
\end{eqnarray}


Following  the approach given in Ref.\cite{Po} we get
\begin{eqnarray}
I_0(t) &=& {\left (\frac{t}{\tau} \right)}^{ \lambda} \,\sum_{n = 0}^{\infty} \,
\frac{{\left(\frac{-\omega^2 \, {t}^{2 + \lambda}}{ {\tau}^{ \lambda}} \right)}^{ n}}{ n!} \sum_{m = 0}^{\infty}
\frac{{\left(-\frac{\bar{\gamma}_{\lambda}\,t^2}{ {\tau}^{ \lambda}
}\right)}^{m}}{ m!}
\nonumber \\
& &    \times \, \, t^{2} \,
E_{\lambda, 3 + 2 n + \lambda +(2 - \lambda) m }^{(n + m)} ( -( t/\tau
)^{\lambda}) \, ,
\label{solukerI0}
\\   \nonumber \\
 I_1(t) &=& \sum_{n = 0}^{\infty} \,
 \frac{{\left(\frac{-\omega^2 \, {t}^{2 + \lambda}}{ {\tau}^{ \lambda}} \right)}^{ n}}{ n!} \sum_{m = 0}^{\infty}
 \frac{{\left(-\frac{\bar{\gamma}_{\lambda}\,t^2}{ {\tau}^{ \lambda}
 }\right)}^{m}}{ m!}
 \nonumber \\
 & &    \times \, \, t^{2} \,
 E_{\lambda, 3 + 2 n + (2 - \lambda) m }^{(n + m)} ( -( t/\tau
 )^{\lambda}) \, ,
\label{solukerI1}
\end{eqnarray}
where $E_{\alpha, \beta} (y)$ is the generalized Mittag-Leffler
function \cite{Go}  defined by the series expansion
\begin{eqnarray}
E_{\alpha, \beta}(y) = \sum_{j = 0}^{\infty}  \frac{y^{j}}{\Gamma(
\alpha j + \beta)}    , \quad \alpha > 0, \quad \beta > 0 \, ,
\label{mittaglef}
\end{eqnarray}
and $E_{\alpha,\beta}^{(k)} (y)$ is the derivative of the Mittag-Leffler
function
\begin{eqnarray}
E_{\alpha,\beta}^{(k)} (y) =  \frac{ d^k}{dy^k}E_{\alpha,\beta}(y) =
\sum_{j = 0}^{\infty} \frac{( j + k )! \, y^{j}}{j ! \,\Gamma(
\alpha ( j + k ) + \beta)} \, . \label{demittaglef}
\end{eqnarray}


Then, from (\ref{kernellades})
\begin{eqnarray}
I(t) =  I_0(t)  + I_1(t) \,   ,
\label{soluI}
\end{eqnarray}
where $ I_0(t)$ and $ I_1(t) $ are given by  (\ref{solukerI0})
and (\ref{solukerI1}) respectively.

The relaxation functions $G(t)$ and  $g(t)$  can be calculated
using (\ref{relaIG}), (\ref{relaGg}) and the relation
\cite{Po}
\begin{eqnarray}
\frac{d}{dt} ( t^{\alpha k + \beta -1} E_{\alpha,\beta}^{(k)}
(-\gamma \, t^{\alpha})) =  t^{\alpha k + \beta -2}
E_{\alpha,\beta -1}^{(k)} (-\gamma \, t^{\alpha}) \, .
\label{reladeg}\nonumber \\
\end{eqnarray}

Then, we get
\begin{eqnarray}
G(t) =  G_0(t)  + G_1(t) \,  ,
\label{soluG}
\end{eqnarray}
where
\begin{eqnarray}
G_0(t) &=& {\left (\frac{t}{\tau} \right)}^{ \lambda} \,\sum_{n = 0}^{\infty} \,
\frac{{\left(\frac{-\omega^2 \, {t}^{2 + \lambda}}{ {\tau}^{ \lambda}} \right)}^{ n}}{ n!} \sum_{m = 0}^{\infty}
\frac{{\left(-\frac{\bar{\gamma}_{\lambda}\,t^2}{ {\tau}^{ \lambda}
}\right)}^{m}}{ m!}
\nonumber \\
& &    \times \, \, t \, \,
E_{\lambda, 2 + 2 n + \lambda +(2 - \lambda) m }^{(n + m)} ( -( t/\tau
)^{\lambda}) \, ,
\label{solukerG0}
\\   \nonumber \\
 G_1(t) &=& \sum_{n = 0}^{\infty} \,
 \frac{{\left(\frac{-\omega^2 \, {t}^{2 + \lambda}}{ {\tau}^{ \lambda}} \right)}^{ n}}{ n!} \sum_{m = 0}^{\infty}
 \frac{{\left(-\frac{\bar{\gamma}_{\lambda}\,t^2}{ {\tau}^{ \lambda}
 }\right)}^{m}}{ m!}
 \nonumber \\
 & &    \times \, \, t \,
 E_{\lambda, 2 + 2 n + (2 - \lambda) m }^{(n + m)} ( -( t/\tau
 )^{\lambda}) \, ,
\label{solukerG1}
\end{eqnarray}
and
\begin{eqnarray}
g(t) =  g_0(t)  + g_1(t)
\label{solug}
\end{eqnarray}
where
\begin{eqnarray}
g_0(t) &=& {\left (\frac{t}{\tau} \right)}^{ \lambda} \,\sum_{n = 0}^{\infty} \,
\frac{{\left(\frac{-\omega^2 \, {t}^{2 + \lambda}}{ {\tau}^{ \lambda}} \right)}^{ n}}{ n!} \sum_{m = 0}^{\infty}
\frac{{\left(-\frac{\bar{\gamma}_{\lambda}\,t^2}{ {\tau}^{ \lambda}
}\right)}^{m}}{ m!}
\nonumber \\
& &    \times \,
E_{\lambda, 1 + 2 n + \lambda +(2 - \lambda) m }^{(n + m)} ( -( t/\tau
)^{\lambda}) \, ,
\label{solukerg0}
\\   \nonumber \\
g_1(t) &=& \sum_{n = 0}^{\infty} \,
\frac{{\left(\frac{-\omega^2 \, {t}^{2 + \lambda}}{ {\tau}^{ \lambda}} \right)}^{ n}}{ n!} \sum_{m = 0}^{\infty}
\frac{{\left(-\frac{\bar{\gamma}_{\lambda}\,t^2}{ {\tau}^{ \lambda}
}\right)}^{m}}{ m!}
\nonumber \\
& &    \times \,
E_{\lambda, 1 + 2 n + (2 - \lambda) m }^{(n + m)} ( -( t/\tau
)^{\lambda}) \, .
\label{solukerg1}
\end{eqnarray}

It is worth mentioning that expressions  (\ref{soluI}),
(\ref{soluG}) and (\ref{solug}) fully determine the temporal
evolution of the mean values (\ref{VMX}) and (\ref{VMV}),
 variances (\ref{sigmaxni}) to (\ref{sigmaxvni}), mean square displacement (\ref{msd})
 and velocity autocorrelation function (\ref{cv}).

Notice that in the limit  $ \omega \to 0 $
only survive the terms with   $ n = 0 $ in  equations
(\ref{solukerI0}) and (\ref{solukerI1}). Then, Eq.(\ref{gammanuevo}) reduces to
$\bar{\gamma}_{\lambda} = \gamma_{\lambda}$,
and the expression of the relaxation function $I(t)$ for the free particle case \cite{VD1} is recovered.


On the other hand, in the limit $ \tau \to 0 $ the function $I_1(t)$ vanishes and
the behavior  of $I_0(t)$ can be achieved introducing
the asymptotic behaviors of the
generalized Mittag-Leffler function \cite{Po}
\begin{eqnarray}
E_{\alpha, \beta}(-y) \thicksim   \frac{1}{ y \, \Gamma( \beta-
\alpha)},  \quad    y > 0, \label{mitaglfpas}
\end{eqnarray}
and its derivative
\begin{eqnarray}
E_{\alpha, \beta}^{(k)}(-y) \thicksim  \frac{k!}{
y^{k+1}} \frac{ 1}{ \Gamma( \beta- \alpha)}
\label{mitaglfpas2}
\end{eqnarray}
in Eq. (\ref{solukerI0}).
Then,  after some algebra we obtain
\begin{eqnarray}
 I(t) & = & \lim_{\tau \to 0} I_0(t) = \sum_{n = 0}^{\infty} \,
\frac{{\left(-\omega^2 \, t^{2} \right)}^{ n}}{n!} \,     \nonumber \\
& &
\times \, t^{2} \, E_{2-\lambda, 3 + \lambda n}^{(n)} (-\gamma_{\lambda}\, t^{2- \lambda}) \, ,
\label{solukerIosclp}
\end{eqnarray}
where we have used that  $\bar{\gamma}_{\lambda} \to  \gamma_{\lambda}$  for $ \tau \to 0 $,
according to (\ref{gammanuevo}).
The expression in series given in Eq.(\ref{solukerIosclp}) coincides with the expression
for the relaxation integral function $I(t)$
corresponding to a pure power-law correlation function,
previously obtained in Ref.\cite{VD}.

Likewise, one can verify that in the limit $ \tau \to 0 $ the
relaxation functions $G(t)$ and  $g(t)$ are also the same to that in the case
of a pure power-law correlation function,   given by
\begin{eqnarray}
G(t) & = &  \sum_{n = 0}^{\infty} \,
\frac{{\left(-\omega^2 \, t^{2} \right)}^{ n}}{n!}
\, t \, E_{2-\lambda, 2 + \lambda n}^{(n)} (-\gamma_{\lambda}\,t^{2- \lambda}) \, ,
\label{solukerGosclp}
\\
g(t) & = &  \sum_{n = 0}^{\infty} \,
 \frac{{\left(-\omega^2 \, t^{2} \right)}^{ n}}{n!}
 \, E_{2-\lambda, 1 + \lambda n}^{(n)} (-\gamma_{\lambda}\, t^{2- \lambda}) \, .
 \label{solukergosclp}
 \end{eqnarray}

\section{Temporal  behavior of the relaxation functions}

The analytical expressions  (\ref{soluI}),
(\ref{soluG}) and (\ref{solug}) are the main result of this work.
In the following we will analyze the time behavior of the relaxation
functions for different regimes.


The short-time behavior ($t\ll\tau$) of the relaxation functions
can be obtained using the series expansions (\ref{mittaglef}) and  (\ref{demittaglef}).
Then
\begin{eqnarray}
I(t) & \approx &  \frac{t^{2}}{2}
- \left(\frac{\gamma_{\lambda}}{\tau^{\lambda}}+\omega^2 \right)\,\frac{t^{4}}{24}
 \, ,
\\
G(t) & \approx &  t
- \left(\frac{\gamma_{\lambda}}{\tau^{\lambda}}+\omega^2 \right) \, \frac{t^{3}}{6}\, ,
\\
g(t)& \approx & 1
- \left(\frac{\gamma_{\lambda}}{\tau^{\lambda}}+\omega^2 \right) \, \frac{t^{2}}{2} \, ,
\end{eqnarray}
which are  the expected for a
harmonic oscillator driven by a  noise  with a finite correlation at the origin \cite{Wang3, Wa,Vi}.


Now we get an expression of the function $I(t)$ for times bigger than
the characteristic time $\tau$ of the noise, i.e.   $  t \gg \tau$.
For this purpose we introduce the approximation (\ref{mitaglfpas2})
in Eqs. (\ref{solukerI0}) and (\ref{solukerI1}).
After some algebra we get
\begin{eqnarray}
\label{I0asin}
I_0(t) & \approx & \sum_{n = 0}^{\infty}  \frac{(-1)^n}{n!}
\, (\omega^{2} t^{2})^{n}
\nonumber \\
& &
\times \, t^{2} \, E_{2-\lambda,  3 + \lambda n}^{(n)} (-\bar{\gamma}_{\lambda}\,t^{2- \lambda}) \, ,
\end{eqnarray}
and
\begin{eqnarray}
\label{I1asin}
 I_1 (t) & \approx & \left( \frac{t}{\tau}\right)^{-\lambda} \sum_{n = 0}^{\infty}  \frac{(-1)^n}{n!}
\, (\omega^{2} t^{2})^{n}
\nonumber \\
& &
\times \, t^{2} \, E_{2-\lambda,  3 -\lambda + \lambda n }^{(n)} (-\bar{\gamma}_{\lambda}\,t^{2- \lambda}) \, .
\end{eqnarray}
%


Let us analyze  the  behaviors of the relaxation functions $I(t)$,
$G(t)$ and $g(t)$ for $
\bar{\gamma}_{\lambda}\,t^{2- \lambda} \gg 1 $.
Introducing the approximation  (\ref{mitaglfpas2}) in (\ref{I0asin}) and (\ref{I1asin}),
after some calculations and using (\ref{mittaglef1p}) one gets
\begin{eqnarray}
I(t) \approx  \frac{1}{\omega^2} \, \left \{ 1  - \nu \, E_{\lambda} \left( -
\nu\frac{\omega^2 }{ \gamma_{\lambda}} \,  t^{\lambda}\right)  \right \} \, .
 \label{kerIdesasin}
\end{eqnarray}

Then, from (\ref{relaIG}) and (\ref{relaGg}) we get
\begin{eqnarray}
G(t) \approx - \frac{\nu} {\omega ^2} \, \frac{d} {dt}E_{\lambda} \left( -
\nu\frac{\omega^2 }{ \gamma_{\lambda}} \,  t^{\lambda}\right) \, ,
 \label{kerdesasin}
\end{eqnarray}
and
\begin{eqnarray}
g(t) \approx - \frac{\nu} {\omega ^2} \, \frac{d^2} {d^2t}E_{\lambda}
\left( -\nu
\frac{\omega^2 }{ \gamma_{\lambda}} \,  t^{\lambda}\right)  \, ,
 \label{kergdesasin}
\end{eqnarray}
where we introduced  the dimensionless factor
\begin{eqnarray}
\nu = \frac{\gamma_{\lambda}}{\bar{\gamma}_{\lambda} } \, ,\qquad 0<\nu\leq 1 \,.
\,\label{nudef}
\end{eqnarray}

The relaxation functions (\ref{kerIdesasin}) to (\ref{kergdesasin}) have the same functional
 form to the  results obtained in the pure power-law case \cite{VD} but  with the presence of the scale factor $\nu$.
In the limit $ \tau \to 0 $ is $\nu=1 $ and one recovers the expressions
 corresponding to a pure power-law  noise \cite{VD}.

It is worth pointing out that these expressions are the same to those that can be obtained
directly discarding
the inertial term $s^{2}$ in (\ref{kernella}).
Then, Eqs. (\ref{kerIdesasin}) to (\ref{kergdesasin}) represents
the solutions in the high friction limit.


The strictly asymptotic behavior of the relaxation functions $I(t)$,  $G(t)$
and $g(t)$  can be obtained introducing the asymptotic behavior
(\ref{mitag1pas}) of the Mittag-Leffler function  in Eqs.
(\ref{kerIdesasin}) to (\ref{kergdesasin}). Then, for
$\nu\,\frac{\omega^2 }{ \gamma_{\lambda}} \,  t^{\lambda}\gg 1 $
the relaxation functions can be  written as
\begin{eqnarray}
 I(t)&\approx& \frac{1} {\omega ^2} - \frac{\gamma_{\lambda}} {\omega ^4} \, \frac{\sin (\lambda
\, \pi ) }{\pi} \, \frac{\Gamma(\lambda) }{ t ^{\lambda }} \, , \label{kerdesasin0}\\
 G(t)& \approx &   \frac{\gamma_{\lambda}} {\omega ^4} \, \frac{\sin (\lambda \,
\pi ) }{\pi} \, \frac{\Gamma(\lambda + 1) }{ t ^{\lambda + 1 }} \,
,  \label{kerdesasin1} \\
g(t) &\approx&  -\frac{\gamma_{\lambda}} {\omega ^4}
\, \frac{\sin (\lambda \, \pi ) }{\pi} \, \frac{\Gamma(\lambda +
2) }{ t ^{\lambda + 2 }} \, .
 \label{kerdesasin2}
\end{eqnarray}

As expected, the relaxation functions (\ref{kerdesasin0}) to (\ref{kerdesasin2}) behave  as
a power law in the long-time limit.
These results are in agreement with those obtained in Refs.\cite{VD,DV}
due to the fact that the Mittag-Lefler noise decays as a power-law for very large times.
In the same way, substitution of these asymptotic expansions into
Eqs. (\ref{sigmaxni}) to (\ref{sigmaxvni}) give the long-time
behavior of the variances of the process, which again coincide with those obtained
in Refs.\cite{VD,DV}.

\section{Conclusions}

In this work we have presented an analytically resoluble model for the dynamics
of a classical  harmonic oscillator in a complex environment, which is valid for
all  time range.
We have shown that an anomalous diffusion
process can be generated by a Mittag-Leffler   noise  deriving
exact expressions for the relaxation functions of the oscillator in terms of
the generalized Mittag-Leffler function and its derivatives.
Moreover, in the appropriate limits  the results for
a harmonic oscillator driven by a  power-law  noise is recovered.
However, differences in relation to the usually employed pure power-law
 noise  appear in the interval of short and intermediate times.
For  times shorter than the characteristic time of the noise
the relaxation functions include  a correction due to the presence
of the characteristic time $\tau$.
In the range of intermediate times, the relaxation functions have a
similar functional form to the previously obtained for a  pure
power-law  noise \cite{VD}, but with the inclusion of a
scaling dimensionless parameter.
Finally, in the strictly asymptotic limit  we recover the
anomalous behavior of an  harmonically bounded particle driven by
a  power-law  noise, which is in agreement with the
previous results given in Refs.\cite{VD,DV}.


\begin{acknowledgments}

This work (Desp\'osito and Vi\~{n}ales) was performed under Grant N$^\circ$  PICT
31980/05 from Agencia Nacional de Promoci\'{o}n Cient\'{i}fica y
Tecnol\'{o}gica, and Grant N$^\circ$ X099 from Universidad de Buenos Aires, Argentina.
K.G. Wang is pleased to acknowledge partial financial support received from the Materials World Network Program and Metallic Materials and Nanostructures Program of National Science Foundation, Washington, DC, under Grants DMR-0710484 and from Florida Solar Energy Center.
K.G. Wang is grateful to Drs. Desp\'osito and Vi\~{n}ales' warm hospitality and fruitful discussions when he visited there in 2008.
\end{acknowledgments}




\end{document}